\documentclass[12pt]{JHEP3}
\usepackage{epsfig}

\title{
On the nature of the finite-temperature transition in QCD
}
\author{Agostino Butti\\
        Scuola Normale Superiore,\\
        Piazza dei Cavalieri 7, I-56126 Pisa, Italy.\\
	E-mail: \email{butti@sns.it} 
} 
\author{Andrea Pelissetto\\ 
	Dipartimento di Fisica dell'Universit\`a 
	di Roma  ``La Sapienza'' and INFN, \\ 
        Piazzale Moro 2, I-00185 Roma, Italy.\\
	E-mail: \email{Andrea.Pelissetto@roma1.infn.it}
}
\author{Ettore Vicari \\ 
	Dipartimento di Fisica dell'Universit\`a 
	di Pisa and INFN, \\
        Via Buonarroti 2, I-56127 Pisa, Italy.\\
	E-mail: \email{vicari@df.unipi.it} 
}
\abstract{
We discuss the nature of the finite-temperature transition in QCD
with $N_f$ massless flavors. Universality arguments show that 
a continuous (second-order) transition must be related to 
a three-dimensional
universality class characterized by a complex
$N_f\times N_f$ matrix order parameter and by the symmetry-breaking
pattern 
$[SU(N_f)_L\otimes SU(N_f)_R]/\mathbb{Z}(N_f)_V\rightarrow 
SU(N_f)_V/\mathbb{Z}(N_f)_V$,
or $[U(N_f)_L\otimes U(N_f)_R]/U(1)_V\rightarrow U(N_f)_V/U(1)_V$ if
the $U(1)_A$ symmetry is effectively restored at $T_c$.
The existence of any of these universality classes
requires the presence of a stable fixed point 
in the corresponding three-dimensional $\Phi^4$ theory 
with the expected symmetry-breaking pattern.
Otherwise, the transition is of first order.

In order to search for stable fixed points in these
$\Phi^4$ theories, we exploit a three-dimensional 
perturbative approach in which physical quantities are 
expanded in powers of appropriate renormalized quartic couplings. 
We compute the corresponding 
Callan-Symanzik $\beta$-functions to six loops.
We also determine the large-order behavior to further constrain the analysis.
No stable fixed point is found, except for $N_f = 2$, corresponding 
to the symmetry-breaking pattern
$[SU(2)_L\otimes SU(2)_R]/\mathbb{Z}(2)_V\rightarrow SU(2)_V/\mathbb{Z}(2)_V$ 
equivalent to $SO(4)\rightarrow SO(3)$.
Our results confirm and put on a firmer ground
earlier analyses performed close to four dimensions, based
on first-order calculations in the framework of the 
$\epsilon\equiv 4-d$ expansion.

These results indicate that
the finite-temperature phase transition in QCD is
of first order for $N_f\geq 3$.
A continuous transition is allowed only for $N_f=2$. 
But, since the theory with 
symmetry-breaking pattern 
$[U(2)_L\otimes U(2)_R]/U(1)\rightarrow U(2)_V/U(1)$ 
does not have stable fixed points,
the transition can be  continuous only if
the effective breaking of the $U(1)_A$ symmetry is sufficiently large. 
}
\begin{document}

\section{Introduction}
\label{intro}

The thermodynamics of quarks and gluons described 
by QCD is characterized by a transition from 
a low-temperature hadronic phase, in which chiral symmetry
is broken, to a high-temperature phase with deconfined quarks and gluons
(quark-gluon plasma), in which chiral symmetry is restored.
The main features of this transition depend crucially on the QCD parameters, 
such as the number $N_f$ of flavors and the quark masses.
In the light-quark regime the nature of the finite-temperature transition
is essentially related to the restoring of the chiral symmetry
\cite{PW-84,Wilczek-92,RW-93,GGP-94}.
For recent reviews see, e.g., 
Refs.~\cite{Rajagopal-95,Wilczek-00,Rischke-03,KL-03}.

%%The QCD Lagrangian is
%%\begin{equation}
%%{\cal L} = {1\over 4} F_a^{\mu\nu} F_a^{\mu\nu} + \sum_{f=1}^{N_f} 
%%\bar{\psi}_f (-i \gamma^\mu D_\mu + m_f) \psi_f  
%%\label{qcdla}
%%\end{equation}
In the vanishing quark-mass limit,
the QCD Lagrangian is invariant under $U(N_f)_L$ and $U(N_f)_R$ transformations 
involving the left- and right-handed quark flavors.
Since 
$U(N)_{L,R} \cong  U(1)_{L,R}\otimes [SU(N)/\mathbb{Z}(N)]_{L,R}$
and the group $U(1)_L\otimes U(1)_R$ is isomorphic
to the group $U(1)_V\otimes U(1)_A$ of vector and axial
$U(1)$ transformations,
the classical symmetry group of the theory can be written as
\begin{equation}
U(1)_V\otimes U(1)_A\otimes [SU(N_f)/\mathbb{Z}(N_f)]_L
                     \otimes [SU(N_f)/\mathbb{Z}(N_f)]_R
\end{equation}
The vector subgroup $U(1)_V$ corresponds to the quark-number conservation,
and it is not expected to play any role in the transition.
The axial subgroup $U(1)_A$ is broken by the anomaly at the quantum level,
reducing the relevant symmetry to \cite{PW-84}
$[SU(N_f)_L\otimes SU(N_f)_R]/\mathbb{Z}(N_f)_V$.
At $T=0$ the symmetry is spontaneously
broken to $SU(N_f)_{V}$ with $N_f^2-1$ Goldstone particles (pions
and kaons) with a nonzero quark condensate $\langle \bar{\psi} \psi \rangle$.
With increasing $T$, a phase transition 
characterized by the restoring of the chiral symmetry
is expected at $T_c$; above $T_c$ the quark condensate vanishes.
The symmetry-breaking pattern at the transition is expected to be
\begin{equation}
[SU(N_f)_L\otimes SU(N_f)_R]/\mathbb{Z}(N_f)_V\rightarrow 
SU(N_f)_V/\mathbb{Z}(N_f)_V,
\label{sym1}
\end{equation}
with order parameter given by the 
expectation value of the quark bilinear 
$\Psi_{ij} \equiv \bar{\psi}_{L,i} \psi_{R,j}$.
In this picture the quark masses act as an external field $H_{ij}$ coupled 
to the order parameter.
Note that, if the $U(1)_A$ symmetry is effectively restored at $T_c$,
the relevant symmetry-breaking pattern would be
\begin{equation}
[U(N_f)_L\otimes U(N_f)_R]/U(1)_V \rightarrow U(N_f)_V/U(1)_V \; .
\label{sym2}
\end{equation}
The effects of the anomaly tend to be
suppressed in the large-$N_c$ limit, where $N_c$ is the number
of colors. The symmetry-breaking pattern (\ref{sym2})
should be recovered in the limit of an infinite number of colors.

In the case of a continuous transition, 
thermodynamic quantities are analytic functions of $T$ as long as 
the external field coupled to the order parameter is nonvanishing.
Therefore, if 
the chiral-symmetry restoring transition is continuous
in the vanishing quark-mass limit, 
an analytic crossover is expected for
nonvanishing quark masses
(actually for not too large values because in the
heavy-quark limit the transition is first order \cite{SY-82}).
On the other hand, if the phase transition is of first order in the chiral
limit,
it persists also for
nonvanishing masses, up to a critical surface in the quark-mass phase diagram
where the transition becomes continuous and is expected to be Ising-like 
\cite{GGP-94}.
Outside this surface, i.e. for larger quark masses,
the phase transition disappears and we have an analytic crossover.

The nature of the finite-temperature transition in QCD with 
$N_f$ light flavors has been much investigated, exploiting
various approaches, including numerical simulations of lattice QCD.
Our understanding is still essentially based on the universality
arguments reported by  Pisarski and Wilczek in Ref.~\cite{PW-84}.
Their main points can be summarized as follows:

\begin{itemize}

\item[(i)]
Let us first assume that the phase transition at $T_c$ is continuous
(second order) for vanishing quark masses.
In this case the critical behavior should be described by an effective
three-dimensional (3-$d$) theory. Indeed,
the length scale of the critical modes diverges approaching $T_c$,
becoming eventually much larger
than $1/T_c$ that is the size of the euclidean ``temporal'' dimension
at $T_c$. 
Therefore, the asymptotic critical behavior must be associated
with a 3-$d$ universality class
characterized by a complex $N_f\times N_f$ matrix order parameter,
corresponding to the bilinear $\Psi_{ij}=\bar{\psi}_{L,i} \psi_{R,j}$,
and by the symmetry-breaking pattern (\ref{sym1}),
or (\ref{sym2}) if the $U(1)_A$ symmetry is effectively restored at $T_c$.

\item[(ii)]
According to renormalization-group (RG)  theory,
the existence of such universality classes can be investigated 
by considering the most general 
Landau-Ginzburg-Wilson (LGW)
$\Phi^4$ theory for a complex $N_f\times N_f$ matrix field $\Phi_{ij}$
with the desired symmetry and symmetry-breaking pattern.
The LGW Lagrangian for the symmetry-breaking pattern 
(\ref{sym2}) is given by
\begin{eqnarray}
{\cal L}_{U(N_f)} = {\rm Tr} (\partial_\mu \Phi^\dagger) (\partial_\mu \Phi)
+r {\rm Tr} \Phi^\dagger \Phi 
+ {u_0\over 4} \left( {\rm Tr} \Phi^\dagger \Phi \right)^2
+ {v_0\over 4} {\rm Tr} \left( \Phi^\dagger \Phi \right)^2 .
\label{genun}
\end{eqnarray}
The chiral anomaly reduces the symmetry to (\ref{sym1}) and thus new terms 
must be added. The most relevant one is proportional to 
$\left( {\rm det} \Phi^\dagger + {\rm det} \Phi \right)$, which is 
a polynomial of order $N_f$ in the field $\Phi$. 
Thus, for $N_f\ge 5$ such a term, and therefore the 
$U(1)_A$-symmetry breaking, is irrelevant at the transition. 
Instead, for $N_f=3$ and $N_f=4$ the new term must be added to the 
Lagrangian (\ref{genun}), obtaining
\begin{equation}
{\cal L}_{SU(N_f)} = {\cal L}_{U(N_f)}
+ w_0 \left( {\rm det} \Phi^\dagger + {\rm det} \Phi \right).
\label{gensun}
\end{equation}
For $N_f \le 2$ additional terms depending on the determinant of 
$\Phi$ should be added, in order to include all terms with at most 
four fields compatible with the symmetry.
Nonvanishing quark masses can be accounted 
for by adding an external-field term $H_{ij}$, such as 
\begin{equation}
{\rm Tr} \left( H \Phi + {\rm h.c.}\right).
\end{equation}

\item[(iii)]
The critical behavior at a continuous transition 
is described by the stable fixed point (FP) of the theory,
which determines the universality class.
The absence of a stable FP indicates that the phase transition
is not continuous.
Therefore, a necessary condition of consistency with the 
initial hypothesis that the transition is continuous, cf.~(i),
is the existence of stable FP's in the theories described by
the Lagrangians (\ref{genun}) and (\ref{gensun}).

\item[(iv)] 
In the one-flavor case, $N_f=1$, 
the Lagrangian ${\cal L}_{U(N_f)}$ is equivalent to the
O(2)-symmetric real $\Phi^4$ Lagrangian,
which has a stable FP describing the XY universality class.
For $N_f=1$ the determinant term in the Lagrangian 
${\cal L}_{SU(N_f)}$ is equivalent to
an external field coupled to the order parameter,
which smooths out the singularity of the continuous
transition.

In the two-flavor case, $N_f=2$, the symmetry-breaking pattern (\ref{sym1})
is equivalent to $SO(4)\rightarrow SO(3)$, which is the 
symmetry-breaking pattern of the standard $SO(4)$-symmetric theory for a 
four-dimensional vector field. As is well known, such a theory has a stable 
FP, corresponding to the SO(4) universality class. 
Thus, for $N_f=2$, there exists a three-dimensional universality class 
with the desired symmetry-breaking pattern.
This fact tells us that, 
if the phase transition  is continuous, it belongs to 
the SO(4) universality class, i.e. it has the same critical exponents,
scaling functions, critical equation of state, etc....
However, these arguments do not exclude that the transition is a 
first-order one.

\item[(v)]
The problem is more complex in the other cases, i.e.
$N_f\geq 3$ for the symmetry breaking (\ref{sym1}) and
$N_f\geq 2$ for the symmetry breaking (\ref{sym2}).
In order to investigate the RG flow of the Lagrangians
(\ref{genun}) and  (\ref{gensun}),
Pisarski and Wilczek \cite{PW-84} performed 
a first-order perturbative
calculation within the $\epsilon$-expansion framework,
i.e. an expansion in powers of $\epsilon\equiv 4-d$.
Within this approximation, they did not find stable FP's.
Then, extending this result to three dimensions, i.e. to $\epsilon=1$,
they argued that the finite-temperature transition
in QCD is always a first-order one, with the only possible exception of 
$N_f=2$.
\end{itemize}

The $\epsilon$ expansion provides useful indications for the 
behavior of the RG flow in three dimensions. 
But the validity of the extrapolation to $\epsilon=1$
of the results obtained near four dimensions is not guaranteed,
not even at a qualitative level, especially
if they are obtained from low-order computations.
The location and the stability of
the FP's may drastically change approaching $d=3$,
and new FP's, not present for $d\approx 4$, 
may appear in three dimensions.  
In several physically interesting cases 
low-order (and in some cases also high-order)
$\epsilon$-expansion calculations  fail to provide
the correct physical picture.
We mention the 3-$d$ O($N$)$\oplus$O($M$) invariant
$\Phi^4$ theory describing the multicritical behavior
arising from the competition of distinct order parameters
\cite{KNF-76,CPV-03},\footnote{
Particularly important cases are $N=1$, $M=2$, relevant 
for anisotropic antiferromagnets, and 
$M=2$, $N=3$ relevant for high-$T_c$ superconductors.}
the 3-$d$ O($N$)$\otimes$O($M$) $\Phi^4$ theory 
describing the critical behavior of frustrated spin models with noncollinear
order \cite{Kawamura-88,PRV-01},\footnote{
These models are physically realized by stacked
triangular antiferromagnets, in the case $N=2$, $M=2,3$.}
and  the Ginzburg-Landau
model of superconductors, in which 
a complex scalar field couples to a gauge field 
\cite{HLM-74,kleinertbook}.
For example, in the latter case one-loop $\epsilon$-expansion  calculations
\cite{HLM-74}
indicate that no stable FP exists unless the number $N$ of real components
of the scalar field is larger than $N_c=365$.
This number is much larger than the physical value $N=2$.
Consequently, a first-order transition was always expected \cite{HLM-74}.
Later, exploiting three-dimensional theoretical approaches 
(see, e.g., Ref.~\cite{kleinertbook}) and
Monte Carlo simulations (see, e.g., Ref.~\cite{MCs}),
it was realized that three-dimensional systems described by the 
Ginzburg-Landau model can also undergo a continuous transition---this
implies the presence of a stable FP in the 3-$d$ Ginzburg-Landau theory---in
agreement with experiments \cite{GN-94}.
Therefore, a three-dimensional analysis, 
not relying on extrapolations from $\epsilon\ll 1$
to $\epsilon=1$, is called for in order to check the
picture reported at point (v) above.

In this paper we return to point (v), i.e.
to the issue concerning the existence  
of three-dimensional universality classes with the symmetry-breaking
pattern of QCD, and therefore the existence of stable FP's 
in the theories defined by the Lagrangians (\ref{genun}) and (\ref{gensun}).
In order to obtain a reliable picture of the RG
flow, we perform a calculation in a 
perturbative framework in which the 
dimension $d$ is fixed to the physical value, i.e. $d=3$,
and the expansion is performed in powers of appropriate
renormalized quartic couplings.
We compute the corresponding Callan-Symanzik
$\beta$-functions to six loops.
Moreover, we study the general properties of the expansions: 
we determine the region in which they are Borel summable and 
compute their large-order behavior to further constrain the analysis.
This allows us to reconstruct the 3-$d$ RG flow,
and in particular to check the existence of FP's,
with a quite high confidence.
We show that no stable FP exists in all cases considered
at  point (v) above,
i.e.  $N_f\geq 3$ for the symmetry breaking (\ref{sym1}) and
$N_f\geq 2$ for the symmetry breaking (\ref{sym2}).
Therefore, we fully confirm the conclusions of Ref.~\cite{PW-84}.
Concerning QCD, these results indicate that
the finite-temperature phase transition is
of first order for $N_f\geq 3$.
A continuous transition is allowed only for $N_f=2$. 
But, since no stable FP is associated with
the symmetry breaking $[U(2)_L\otimes U(2)_R]/U(1)_V\rightarrow U(2)_V/U(1)_V$,
the transition can be  continuous only if
the effective breaking of the $U(1)_A$ symmetry is 
sufficiently large. Otherwise, it is first order.
For a specific effective breaking of
the $U(1)_A$ symmetry, we may also have a mean-field critical behavior
with logarithmic corrections, 
which is associated with the tricritical point that separates 
the first-order and the second-order transition
lines in the $T-g$ plane, where $g$ parametrizes the effective
breaking of the $U(1)_A$ symmetry.

Direct information on the QCD phase transitions 
has been provided by numerical Monte Carlo studies
within the lattice formulation of QCD, see, e.g., the recent
review \cite{KL-03}.
The Monte Carlo results obtained so far are consistent with 
and substantially support the
above-reported three-dimensional RG results.
In the case of three-flavor QCD the evidence 
that the transition is first order is rather 
clear, see, e.g., Refs. \cite{KLS-01,IKSY-96,Brownetal-90,KS-88,GPS-87}.
Moreover, it has been verified that the first-order transition
persists for nonvanishing quark masses up to
Ising-like endpoints \cite{KLS-01}.
In the two-flavor case recent lattice simulations have provided a rather convincing
evidence of a continuous transition at $T_c\approx 172$ MeV,
see, e.g., Refs.~\cite{CP-PACS-01,KLP-01,EHMS-01,KS-01,IKKY-97}, 
whose scaling properties are substantially
consistent with those predicted by the 3-$d$
SO(4) universality. However, a
conclusive evidence in favor of an SO(4) scaling
behavior in the continuum limit has not been achieved yet.
Apparently, a mean-field critical behavior has not been
ruled out yet.

The paper is organized as follows.
In Sec.~\ref{LGW} we  present the general three-dimensional perturbative
framework.
In Sec.~\ref{unsy} we report our calculations
for a theory with symmetry $[U(N)_L\otimes U(N)_R]/U(1)_V$, i.e.
for the Lagrangian (\ref{genun}).\footnote{For simplicity,
from now on we simply write $N$ instead of $N_f$.}
The analysis of the perturbative expansions, using also information
on their large-order behavior, shows
no evidence of stable FP's, for any $N$.
In Sec.~\ref{su4sy} we consider the more general 
Lagrangian (\ref{gensun}), which is symmetric under the smaller
group $[SU(N)_L\otimes SU(N)_R]/\mathbb{Z}(N)_V$.
We present a six-loop analysis for $N=4$, which is
the only case for which no convincing arguments 
exist in favor or against the existence of
a stable FP. Again, our analysis does not find
evidence of a stable FP.
Finally, in the Appendix we discuss in detail the phase diagram 
of the general $\Phi^4$ model
realizing the symmetry-breaking pattern (\ref{sym1})
for $N=2$ with a complex $2\times 2$ matrix order parameter,
which is 
relevant for the finite-temperature transition in two-flavor QCD.

\section{Perturbative expansion in Landau-Ginzburg-Wilson $\Phi^4$ theories} 
\label{LGW}

\subsection{General $\Phi^4$ theories}
\label{GT}

In the RG approach to critical phenomena, many phase transitions can be 
investigated by considering effective LGW theories,
containing up to fourth-order powers of the field components.
The general LGW Lagrangian for an $N$-component field $\phi_i$
can be written as
\begin{equation}
{\cal L}_{\rm LGW} = \int d^d x \Bigl[ 
{1\over 2} \sum_i (\partial_\mu \phi_{i})^2 + 
{1\over 2} \sum_i r_i \phi_{i}^2  + 
{1\over 4!} \sum_{ijkl} u_{ijkl} \; \phi_i\phi_j\phi_k\phi_l \Bigr], 
\label{generalH}
\end{equation}
where the number of independent parameters $r_i$ and $u_{ijkl}$ 
depends on the symmetry group of the theory. An interesting 
class of models are those in which $\sum_i \phi^2_i$ is the 
only quadratic invariant polynomial.
In this case, all $r_i$ are equal, $r_i = r$, and 
$u_{ijkl}$ satisfies the trace condition~\cite{BLZ-76}
\begin{equation}
\sum_i u_{iikl} \propto \delta_{kl}.
\label{traceco}
\end{equation}
In these models, criticality is driven by tuning the single parameter
$r$. Therefore, they describe critical phenomena characterized 
by one (parity-symmetric) relevant parameter, 
which often corresponds to the temperature. Of course, there is also
(at least one) parity-odd relevant parameter that corresponds to 
a term $\sum_i h_i \phi_i$ that can be added to the Hamiltonian
(\ref{generalH}). For symmetry reasons, criticality occurs
for $h_i\to 0$. 
The simplest example of this class of models 
is the O($N$)-symmetric $\Phi^4$ theory, which has
only one quartic parameter $u$:
\begin{equation}
u_{ijkl} = 
{u\over 3}\left(\delta_{ij}\delta_{kl} + \delta_{ik}\delta_{jl} +
                \delta_{il}\delta_{jk} \right).
\label{ONu}
\end{equation}
This model describes several interesting universality classes:
the Ising one for $N=1$ 
(e.g., liquid-vapor transition, transitions in
uniaxial magnetic systems, etc...), 
the XY one for $N=2$
(e.g., superfluid transition in $^4$He, 
transitions in magnets with easy-plane anisotropy, etc...),
the Heisenberg one for $N=3$
(e.g., Curie transition in isotropic magnets);
for  $N\rightarrow 0$ it describes the behavior of dilute 
homopolymers for large degree of polymerization.
See, e.g., Refs.~\cite{ZJbook,PV-r} for recent reviews.
But there are  also other physically interesting 
transitions that are
characterized by more complex symmetries:
for instance, the magnetic transitions in
randomly dilute magnetic systems,
in stacked triangular antiferromagnets,
in Mott insulators, and several 
structural phase transitions.
See, e.g., Refs.~\cite{Aharony-76,PV-r,th2002,Sachdev-03} for reviews.
In this case, the corresponding LGW Lagrangian contains two or more 
quartic couplings, but still only one quadratic invariant.
More general LGW Lagrangians, that allow for the presence
of independent quadratic parameters $r_i$, have been considered
to study the multicritical behavior that 
arises from the competition of distinct types of ordering,
see, e.g., Refs.~\cite{KNF-76,CPV-03}.

In the field-theory (FT) approach the RG flow 
is determined by a set of RG
equations for the correlation functions of the order parameter.
In the case of a continuous transition, 
the critical behavior is determined by the stable FP
of the theory, which characterizes a universality class. 
The absence of a stable FP is usually considered as
an indication that the phase transition is of first order,
even in those cases in which the mean-field approximation predicts
a continuous transition.
But, even in the presence of a stable FP,
a first-order transition may  still occur
for systems that are outside its attraction domain.
 
The RG flow can be studied by perturbative methods, for example 
by performing an expansion in powers of $\epsilon\equiv 4-d$ or a 
fixed-dimension (FD) expansion in powers of appropriate zero-momentum
renormalized quartic couplings. 
The $\epsilon$ expansion, which was introduced by Wilson and Fisher \cite{WF-72}, 
is based on the observation that, for $d=4$, 
the theory is essentially Gaussian. One  considers 
the standard perturbative expansion, and then transforms it
into an expansion in powers of $\epsilon\equiv 4 - d$. 
Subsequently, 
Parisi \cite{Parisi-80} pointed out the possibility of using 
perturbation theory directly at the physical dimensions $d=3$ and $d=2$
in the massive (high-temperature) phase.
In the following we will follow the latter approach.

\subsection{The fixed-dimension expansion}
\label{FDexp}

Let us consider the generic $\Phi^4$ Lagrangian (\ref{generalH})
for an $N$-component real field $\phi_i$,
with a single parameter $r$, i.e. $r_i=r$ for all $r$, so that the 
quartic terms satisfy the trace condition (\ref{traceco}).
In the framework of the FD expansion one works directly in the dimension 
of interest, $d=3$ or $d=2$. 
In this case  the theory is super-renormalizable, since the 
number of primitively divergent diagrams is finite. 
One may regularize the corresponding  integrals by keeping $d$ arbitrary 
and performing an expansion in $\varepsilon=3-d$ or $\varepsilon=2-d$. 
Poles in $\varepsilon$ appear in divergent diagrams. Such divergences
are related to the necessity of performing an infinite renormalization of the 
parameter $r$ appearing in the bare Lagrangian, see, e.g., the discussion
in Ref.~\cite{BB-85}. This problem
can be avoided by replacing $r$ with the mass $m$ defined by 
\begin{equation}
m^{-2} = \, {1\over \Gamma^{(2)}(0)} \, 
     \left. {\partial \Gamma^{(2)}(p^2) \over \partial p^2}\right|_{p^2=0},
\end{equation}
where the function $ \Gamma^{(2)}(p^2)$ is related 
to the one-particle irreducible two-point function  by
\begin{equation}
\Gamma^{(2)}_{ij} (p) = \delta_{ij} \Gamma^{(2)}(p^2).
\end{equation}
Perturbation theory in terms of $m$ and $u_{ijkl}$ is finite. 
The critical limit is obtained for 
$m\to 0$. To handle it, one considers appropriate 
RG functions. Specifically, one defines
the renormalized four-point couplings $g_{ijkl}$ and the
field-renormalization constant $Z_\phi$ by
\begin{eqnarray}
&& 
\Gamma^{(2)}_{ij}(p) = \delta_{ij} Z_\phi^{-1} \left[ m^2+p^2+O(p^4)\right],
\label{ren1g}  \\
&& 
\Gamma^{(4)}_{ijkl}(0) = m^{4-d}\,Z_\phi^{-2} \,g_{ijkl},
\label{rencond}  
\end{eqnarray} 
where $\Gamma^{(n)}_{a_1,\ldots,a_n}$ are $n$-point 
one-particle irreducible correlation functions. 
Eqs.~(\ref{ren1g}) and (\ref{rencond}) relate the mass $m$ and the renormalized
quartic couplings $g_{ijkl}$ to the corresponding bare Lagrangian parameters
$r$ and $u_{ijkl}$.
In addition, one introduces the function $Z_t$ that is defined by the relation
$\Gamma^{(1,2)}_{ij}(0) = \delta_{ij} Z_t^{-1}$,
where $\Gamma^{(1,2)}$ is the one-particle irreducible
two-point function with an insertion of $\frac{1}{2}\sum_i \phi_i^2$.

The common zeroes of the Callan-Symanzik $\beta$-functions
\begin{equation}
\beta_{ijkl}(g_{abcd}) =  
m \left. {\partial g_{ijkl}\over \partial m}\right|_{u_{abcd}} 
\label{bijkl}
\end{equation}
provide the FP's of the theory.
In the case of a continuous transition,
when $m\rightarrow 0$, the couplings $g_{ijkl}$ are driven towards an
infrared-stable FP.
The stability properties of the FP's are 
controlled  by the  eigenvalues $\omega_i$ of the matrix 
\begin{equation}
\Omega_{ijkl,abcd} = { \partial \beta_{ijkl} \over \partial g_{abcd} } 
\end{equation}
computed at the given FP:
a FP is stable if all eigenvalues $\omega_i$  have positive real part.
The critical exponents are then obtained by evaluating 
the RG functions
\begin{eqnarray} 
\eta_\phi(g_{ijkl}) = \left. {\partial \ln Z_\phi\over \partial \ln m}
      \right|_{u} , \qquad
\eta_t(g_{ijkl}) = \left. {\partial \ln Z_t \over \partial \ln m} 
      \right|_u
\label{etapt}
\end{eqnarray}
at the stable FP $g_{ijkl}^*$: 
$\eta = \eta_\phi(g_{ijkl}^*)$, and
$\nu  = [ 2 - \eta_\phi(g_{ijkl}^*) + \eta_t(g_{ijkl}^*)]^{-1}$.
From the perturbative expansion of the correlation functions
$\Gamma^{(2)}$, $\Gamma^{(4)}$, and $\Gamma^{(1,2)}$ and 
using the above-reported relations, one can derive the expansion of the
RG functions $\beta_{ijkl}$, $\eta_\phi$, and $\eta_t$
in powers of $g_{ijkl}$. 

\subsection{Resummation of the series}
\label{FDres}

FT perturbative expansions are divergent. Thus,
in order to obtain accurate results, an appropriate resummation 
is required. 
In the case of the O($N$)-symmetric models, 
accurate results have been obtained from the analysis of the available
FD six-loop series  \cite{BNGM-77} by exploiting Borel summability
(proved in dimension $d<4$ \cite{borelsum})
and the knowledge of the large-order behavior of the expansion,
see, e.g., Refs.~\cite{LZ-77,GZ-98}.
If we consider a quantity $S(g)$ that has a perturbative 
expansion\footnote{
Here the coupling $g$ is normalized
according to Eq.~(\ref{rencond}) with 
$g_{ijkl}= 
{1\over 3}\left(\delta_{ij}\delta_{kl} + \delta_{ik}\delta_{jl} +
                \delta_{il}\delta_{jk} \right)\,g$.
}
\begin{equation}
S(g) \approx \sum s_k g^k,
\end{equation}
the large-order behavior of the coefficients is generally given by
\begin{equation}
s_k \sim k! \,(-a)^{k}\,k^b\,\left[ 1 + O(k^{-1})\right].
\label{lobh}
\end{equation}
Borel summability of the series for $g>0$ requires $a>0$. 
The value of the constant $a$ 
is independent of the particular quantity considered,
unlike the constant $b$.
They can be determined by
means of a steepest-descent calculation in which
the saddle point is a finite-energy solution (instanton)
of the classical field equations with negative 
coupling~\cite{Lipatov-77,BLZ-77}, 
see also Refs.~\cite{ZJbook,Parisibook}.\footnote{
\label{fc} More precisely, the saddle-point $f_s(x)$ is a solution of the 
equation $\Delta f_s = f_s - f_s^3$.
One can show that $a$ is related to its 
classical action, indeed $a=-1/S_f$ where
$S_f=-{3\over 2} \int d^d x \,f_s^4$ \cite{ZJbook}.
} 
In three dimensions
and for any $N$, $a=0.00881962...$, see, e.g., Ref.~\cite{ZJbook}. 
In order to resum the perturbative series, 
one may introduce the Borel-Leroy transform $B(t)$ of $S(g)$, 
\begin{equation}
S(g) = \int_0^\infty t^c e^{-t} B(t),
\label{defSg-Bt}
\end{equation}
where $c$ is an arbitrary number. 
Its expansion is given by
\begin{equation} 
B_{\rm exp}(t) = \sum_k {s_k\over \Gamma(k + c + 1)} t^k.
\label{Bexpansion}
\end{equation}
The constant $a$ that characterizes the large-order behavior 
of the original series is related to the
singularity $t_s$ of the Borel transform $B(t)$ that is nearest 
to the origin: $t_s=-1/a$.  
The series $B_{\rm exp} (t)$ is convergent 
in the disk $|t| < |t_s| = 1/a$ of the complex plane, and also on the 
boundary if $c >  b$. In this domain, one can  compute 
$B(t)$ using $B_{\rm exp} (t)$. However, in order to compute the 
integral (\ref{defSg-Bt}),
one needs $B(t)$ for all positive values of $t$. 
It is thus necessary to perform an analytic continuation of $B_{\rm exp} (t)$. 
For this purpose one may use 
Pad\'e approximants to the series (\ref{Bexpansion}) \cite{BNGM-77}:
This is the so-called Pad\'e-Borel resummation method.
A more refined procedure exploits the knowledge of the large-order
behavior of the expansion, and in particular of the constant $a$ in Eq.~(\ref{lobh}).
One performs an Euler transformation \cite{LZ-77}
\begin{equation}
y(t) = {\sqrt{1 + a t} - 1\over \sqrt{1 + a t} + 1 },
\label{cfmap}
\end{equation}
that allows to rewrite $B(t)$ in the form
\begin{equation}
B(t) = \sum_k f_k \, [y(t)]^k.
\label{Borel-2}
\end{equation}
If all the singularities of $B(t)$ belong to the real interval $[-\infty,-t_c]$,
the expansion (\ref{Borel-2}) converges everywhere in the complex 
$t$-plane except on the negative axis for $t < - t_c$. 
After these transformations, one obtains a new expansion for the 
function $S(g)$:
\begin{equation}
S(g) \approx  \int^\infty_0 dt\, e^{-t} t^c\, \sum_k f_k\,  [y(tg)]^k.
\label{Borel-final}
\end{equation}
This sequence of operations has transformed the original divergent series
into an integral of a convergent one, which can then be studied 
numerically.
However, the convergence of the integral (\ref{Borel-final}), that is 
controlled by the analytic properties of $S(g)$,
is not guaranteed. Indeed, 
since 
generic RG quantities have a cut for $g\ge g^*$ 
\cite{Parisi-80,Nickel-82,PV-98},
where $g^*$ is the FP,
the integral does not converge for $g>g^*$.

The method can be extended to the case
of more general LGW theories, with several quartic couplings,
see, e.g., Refs.~\cite{CPV-00,PV-r}.   
In the case of $k$ independent quartic Lagrangian terms, a generic quantity
$S$ depends on $k$ renormalized couplings. Its expansion 
can be written as
\begin{equation}
S(g_1,...,g_k) = \sum_{j_1,...,j_k} c_{j_1,...,j_k} g_1^{j_1}...g_k^{j_k} =
\sum_n s_n g^n,
\end{equation}
where 
\begin{equation}
s_n = \sum_{j_1+...+j_k=n} c_{j_1,...,j_k} 
\hat{g}_1^{j_1}...\hat{g}_k^{j_k}, \qquad \hat{g}_j = g_j/g.
\end{equation}
The large-order behavior of the coefficients $s_n$,
and therefore the singularities of the Borel transform, 
depends on the ratios $\hat{g}_i$, but it can be still determined by 
steepest-descent calculations.
Borel summability requires that the Borel trasform
does not have singularities on the positive real axis.
Under the assumption that the group and the 
space structure of the relevant saddle points are 
decoupled,\footnote{
This fact has been proved to hold in  
O($N$)-symmetric models \cite{BLZ-77}.}
it is possible to show that the region 
where Borel summability holds is directly related to 
the region of stability of the Lagrangian potential
with respect to the bare quartic couplings: 
with the normalizations implicit in Eq.~(\ref{rencond}), it is enough
to replace the bare quartic couplings with their renormalized counterparts.
For instance, in O($N$)-symmetric models the potential is stable only
for $u > 0$ and therefore the expansion is Borel summable only for $g>0$.
Once the large-order behavior of the coefficients $s_n$ has 
been determined, i.e.
\begin{equation}
s_n \sim n! \,[-A(\hat{g}_1,...,\hat{g}_k)]^n  \, n^b,
\label{large-order}
\end{equation}
one may use a conformal-mapping method based on Eq.~(\ref{cfmap}) 
to resum the series in $g$, as in the O($N$)-symmetric case.
Again, $A(\hat{g}_1,...,\hat{g}_k)$ does not depend on the quantity at hand.
Alternatively, one may use the Pad\'e-Borel method,
employing Pad\'e approximants to analytically extend the Borel transform.

\section{The $U(N)_L\otimes U(N)_R$ effective theory}
\label{unsy}

\subsection{The effective  LGW $\Phi^4$ theory}
\label{LGWun}

The most general $\Phi^4$ Lagrangian that is 
symmetric under  $U(N)_L\otimes U(N)_R$ transformations is given by
\begin{equation}
{\cal L}_{U(N)} = {\rm Tr} (\partial_\mu \Phi^\dagger) (\partial_\mu \Phi)
+r {\rm Tr}\ \Phi^\dagger \Phi 
+ {u_0\over 4} \left( {\rm Tr}\ \Phi^\dagger \Phi \right)^2
+ {v_0\over 4} {\rm Tr} \left( \Phi^\dagger \Phi \right)^2,
\label{un}
\end{equation}
where the fundamental field $\Phi_{ij}$ is a complex $N \times N$
matrix. There is only one quadratic invariant and therefore
the trace condition (\ref{traceco}) is satisfied by the quartic terms. 
The Lagrangian (\ref{un}) is invariant under the transformations
$\Phi \to A \Phi B$, $A$, $B\in U(N)$, and the corresponding symmetry group is 
$[U(N)_L\otimes U(N)_R]/U(1)_V$. 
The condition $v_0>0$ is necessary to realize the symmetry breaking 
\begin{equation}
[U(N)_L\otimes U(N)_R]/U(1)_V \rightarrow U(N)_V/U(1)_V,
\label{symbr}
\end{equation}
where $U(N)_V$ is the group of transformations 
$\Phi \to A \Phi A^\dagger$, $A\in U(N)$.
Indeed, for $r<0$ and $v_0>0$ the minimum of the quartic potential,
which has the form $\Phi_{ij} =\phi_0 \delta_{ij}$, 
is symmetric under $U(N)$ transformations,
thus leading to the desired symmetry-breaking pattern.
For $v_0<0$ the minimum of the potential is obtained for
$\Phi_{ij} = \phi_0 \delta_{ia}\delta_{ja}$,
where $a$ is an arbitrary integer,
$1\leq a \leq N$.  As one can easily check, this leads to a different
symmetry-breaking pattern.
Note that for $N=1$ the two quartic
terms are equal, and the Lagrangian ${\cal L}_{U(N)}$ becomes equivalent
to the standard $U(1)$-symmetric $\Phi^4$ Lagrangian for a complex field 
with coupling $u_0+v_0$, which in turn is directly related to the 
O(2)-symmetric real $\Phi^4$ Lagrangian.

As explained in Sec.~\ref{FDexp}, 
the theory is renormalized by introducing
a set of zero-momentum conditions 
for the one-particle irreducible two-point and four-point 
correlation functions:
\begin{eqnarray}
&&\Gamma^{(2)}_{a_1a_2,b_1b_2}(p) = {\delta^2 \Gamma \over \delta \Phi_{a_1a_2}^\dagger 
\delta \Phi_{b_1b_2}} = \delta_{a_1b_1}\delta_{a_2b_2} Z_\phi^{-1} \left[ m^2+p^2+O(p^4)\right],
\label{ren1}  \\
&&\Gamma^{(4)}_{a_1a_2,b_1b_2,c_1c_2,d_1d_2}(0) =
\left. {\delta^4 \Gamma \over \delta \Phi_{a_1a_2}^\dagger 
\delta \Phi_{b_1b_2}^\dagger \delta \Phi_{c_1c_2} \delta 
\Phi_{d_1d_2}}\right|_{\rm zero \; mom.} =
\label{ren2}  \\
&&\qquad
= Z_\phi^{-2} m^{4-d} \left( u U_{a_1a_2,b_1b_2,c_1c_2,d_1d_2} +  
v V_{a_1a_2,b_1b_2,c_1c_2,d_1d_2}\right)
\nonumber 
\end{eqnarray}
where $\Gamma$ is the generator of the one-particle irreducible correlation functions
(effective action), and
\begin{eqnarray}
&&U_{a_1a_2,b_1b_2,c_1c_2,d_1d_2}=
{1\over 2} \left( \delta_{a_1c_1}\delta_{b_1d_1}\delta_{a_2c_2}\delta_{b_2d_2}+
\delta_{a_1d_1}\delta_{b_1c_1}\delta_{a_2d_2}\delta_{b_2c_2} \right), 
\nonumber \\
&&V_{a_1a_2,b_1b_2,c_1c_2,d_1d_2}=
{1\over 2} \left( \delta_{a_1c_1}\delta_{b_1d_1}\delta_{a_2d_2}\delta_{b_2c_2}+
\delta_{a_1d_1}\delta_{b_1c_1}\delta_{a_2c_2}\delta_{b_2d_2} \right). 
\nonumber
\end{eqnarray}
The FP's of the theory are given by the common zeroes of the
Callan-Symanzik $\beta$-functions
\begin{equation}
\beta_u(u,v) = \left. m{\partial u\over \partial m}\right|_{u_0,v_0},\qquad
\beta_v(u,v) = \left. m{\partial v\over \partial m}\right|_{u_0,v_0}.
\label{betaf}
\end{equation}
Their stability is controlled  by the eigenvalues of the
matrix
\begin{equation}
\Omega =
\left(\matrix{
{\partial \beta_u/\partial u} & {\partial \beta_u/\partial v} 
\cr
{\partial \beta_v/\partial u} & {\partial \beta_v/\partial v} 
}\right) \; .
\label{omegaun}
\end{equation}
A FP is stable if 
all the eigenvalues of its stability matrix have positive real part.
According to RG theory, a stable FP in the region $v>0$ would imply
the existence of a universality class describing
continuous transitions characterized by the symmetry breaking (\ref{symbr}).
Therefore, in order to establish the existence of a 3-$d$
universality class with symmetry-breaking pattern
(\ref{symbr}) and a complex $N\times N$ matrix order parameter, 
one must search for stable zeroes of the $\beta$-functions.

One can easily identify two FP's in the theory described
by the Lagrangian ${\cal L}_{U(N)}$,
without performing any calculation.
The first one is the Gaussian FP for $u=v=0$,
which is always unstable.
Since for $v_0=0$ the Lagrangian ${\cal L}_{U(N)}$ becomes
equivalent to the one of the O($2N^2$)-symmetric model,
the corresponding O($2N^2$) FP must exist in the $v=0$ axis
for $u>0$.
One-loop $\epsilon$-expansion computations do not find
other FP's \cite{PW-84}. 
The location of the O($2N^2$) FP can be obtained by using known 
results for the O($M$)-symmetric theories.
Estimates relevant for our study, i.e. for O(8),
O(18), O(32) and O(72), corresponding to $N=2,3,4,6$,
can be found in Refs.~\cite{AS-95,CPRV-96}.
Moreover, using the results of Refs.~\cite{CPV-03,th2002}
on the stability of the O($N$)-symmetric FP
under generic perturbations, one can already conclude
that the O($2N^2$) FP is unstable for any $N\geq 2$.
Indeed, the term  
${\rm Tr} \left( \Phi^\dagger \Phi \right)^2$ in the Lagrangian
${\cal L}_{U(N)}$
is a particular combination of quartic operators transforming as
the spin-0 and spin-4 representations of the O($2N^2$) group,
and any spin-4 quartic perturbation is relevant
at the O($M$) FP for $M\geq 3$ \cite{CPV-03,th2002}.
The RG dimension $y_{4,4}$ of the spin-4 quartic operators at the O($M$) FP
can be computed by using the
six-loop FD results and the five-loop $\epsilon$-expansion results for the 
related crossover exponent $\phi_{4,4}= y_{4,4}\nu$ reported in 
Ref.~\cite{CPV-03}. We obtain
the estimates $y_{4,4} = 0.386(7), 0.673(6), 0.808(5), 0.914(3)$ 
respectively for
O(8), O(16), O(32), and O(72), that are the cases interesting for this work.
Since $y_{4,4}>0$, the O($2N^2$) FP is unstable in all cases.

\subsection{Fixed-dimension expansion in three dimensions}
\label{fdexpun}

In order to investigate the existence of new FP's, 
we computed the FD expansion of the 3-$d$
Callan-Symanzik  $\beta$-functions (\ref{betaf}) to six loops.
This is rather cumbersome, since approximately one thousand
Feynman diagrams must be evaluated.
We employed a symbolic manipulation program, 
which  generated the diagrams 
and computed the symmetry and group factors of each of them.
We used the numerical results compiled in Ref.~\cite{NMB-77}
for the integrals associated with each diagram.

We report the series in terms of the rescaled couplings $\bar u$ and $\bar v$,
\begin{equation}
u \equiv  {16 \pi \over 4+N^2} \bar u, \qquad 
v \equiv   {16 \pi\over 4+N^2} \bar v .
\label{rescv}
\end{equation}
The six-loop expansions of the $\beta$-functions 
corresponding to these rescaled variables are given by
\begin{eqnarray}
&&\beta_{\bar u}(\bar u,\bar v) = -\bar u
+ \bar u^2 + \frac{4N}{4+N^2} \bar u \bar v + \frac{3}{4+N^2} \bar v^2
-\frac{190+82N^2}{27(4+N^2)^2} \bar u^3 
\label{bu} \\
&&- \frac{400N}{27(4+N^2)^2} \bar u^2 \bar v 
-\frac{370+46N^2}{27(4+N^2)^2} \bar u \bar v^2 -\frac{4N}{(4+N^2)^2} \bar v^3
+ \sum_{i+j \geq 4} \, b^{u}_{ij} \frac{1}{(4+N^2)^{i+j}} \, \bar u^i \bar v^j, 
\nonumber 
\end{eqnarray}
\begin{eqnarray}
&&\beta_{\bar v}(\bar u,\bar v) = -\bar v 
+\frac{6}{4+N^2} \bar u \bar v + \frac{2N}{4+N^2} \bar v^2
-\frac{370+46N^2}{27(4+N^2)^2} \bar u^2 \bar v 
\label{bv} \\
&&- \frac{400N}{27(4+N^2)^2} \bar u \bar v^2
-\frac{136+28N^2}{27(4+N^2)^2} \bar v^3
+ \sum_{i+j \geq 4} \, b^{v}_{ij}\frac{1}{(4+N^2)^{i+j}} \, \bar u^i \bar v^j .
\nonumber
\end{eqnarray}
The coefficients 
$b^{u}_{ij}$ and $b^{v}_{ij}$, with $4\leq i+j\leq  7$, 
are reported in the Tables~\ref{betau} and \ref{betav}, 
respectively.
Actually, we also computed the RG functions $\eta_\phi$ and $\eta_t$,
cf. Eq.~(\ref{etapt}), to six loops. But,
since we will not use them, we do not report their series.
Of course, they are available on request.

We have done a number of checks by using known
results available in the literature.
For $\bar{v}=0$, the expansion of $\beta_{\bar u}$ in powers of $\bar{u}$
reproduces that of the
$\beta$-function of the O($2 N^2$)-symmetric model that can be found 
in Refs.~\cite{NMB-77,AS-95} to six loops.
Since we computed  the series for generic $U(N)_L\otimes U(M)_R$ models,
we also verified nontrivial relations holding for $M=1$ and any $N$, 
such as
\begin{equation}
\beta_u(z+y, z-y;N,M=1) + \beta_v(z+y, z-y;N,M=1) = \bar{\beta}(z;2N),
\end{equation}
where $\bar{\beta}(z;2N)$ is the $\beta$-function of the
O($2N$) model and we are considering unrescaled couplings normalized as in 
Eq.~(\ref{rencond}).

In order to resum the perturbative expansions, 
we use the conformal-mapping method described in Sec.~\ref{FDres}.
For this purpose we compute the function $A$ defined in Eq.~(\ref{large-order})
by studying the saddle-point solutions (instantons)
of the classical field equations.
Under the assumption that the group and space structure of the saddle-point
solutions decouple, the computation 
can be easily reduced to the case of the one-component $\Phi^4$ theory 
described in detail in, e.g.,  Refs.~\cite{ZJbook,Parisibook}.
Here we only report the results, without giving
the details of the calculations which are rather standard.
Consider a generic quantity 
$S(\bar{u},\bar{v})= \sum_{ij} c_{ij} \bar{u}^i \bar{v}^j$ and 
the corresponding expansion 
\begin{equation}
S(g\bar u,g\bar v) = \sum_n s_n(\bar u,\bar v) g^n.
\end{equation}
The large-order behavior is given by
\begin{eqnarray}
&&s_n \sim n! [-A(\bar u,\bar v)]^n \, n^b,
\label{bsing}\\
&&A(\bar u, \bar{v}) = A_N
{\rm Max} \left[ \bar u+\bar v,\, \bar u+\frac{\bar v}{N} \right],
\label{Ares} \\
&& A_N = {24 \pi\over 4 + N^2} a,
\end{eqnarray}
where $a\approx0.00881962$.  The expansion is Borel summable in the region
\begin{equation}
\bar{u} + {\bar{v}\over N} \geq 0,\qquad
\bar{u} + \bar{v} \geq 0,
\end{equation}
which, as already mentioned in Sec.~\ref{FDres},
is related to the stability region of the quartic potential
in the Lagrangian ${\cal L}_{U(N)}$, which is given by
$u_0+v_0/N\geq 0$ and $u_0+v_0\geq 0$ \cite{PW-84}.

\TABLE[ht]{
\footnotesize
\caption{The coefficients $b^{u}_{ij}$, cf. Eq.~(\ref{bu}).  
}
\label{betau}
\begin{tabular}{cl}
\hline\hline
\multicolumn{1}{c}{$i,j$}&
\multicolumn{1}{c}{$b^u_{ij}$}\\
\hline \hline
4,0   &   $99.8202086 + 79.8954292\,N^2 + 16.4329798\,N^4 + 0.674471379\,N^6$ \\
3,1   &   $329.227699\,N + 108.894438\,N^3 + 6.64687835\,N^5$ \\
2,2   &   $366.245988 + 251.386542\,N^2 + 41.2000397\,N^4 + 0.310944602\,N^6$ \\
1,3   &   $280.991858\,N + 83.3331935\,N^3 + 3.27130728\,N^5 $\\
0,4   &   $65.0697173 + 40.0695380\,N^2 + 5.95052716\,N^4 $\\
\hline
5,0   &   $-458.051683 - 415.773535\,N^2 - 111.135357\,N^4 - 8.64375912\,N^6 + 0.0778229492\,N^8 $\\
4,1   &   $-2064.17070\,N - 880.190454\,N^3 - 85.6254168\,N^5 + 1.35288201\,N^7$\\
3,2   &   $-2647.21638 - 2601.32272\,N^2 - 492.107570\,N^4 - 1.80697817\,N^6 $\\
2,3   &   $-3634.53973\,N - 1183.42261\,N^3 - 70.4182766\,N^5 - 0.430339129\,N^7 $\\
1,4   &   $-1349.56809 - 1084.80505\,N^2 - 197.491458\,N^4 - 2.65955034\,N^6 $\\
0,5   &   $-336.753422\,N - 103.463574\,N^3 - 4.81880464\,N^5 $\\
\hline
6,0   &   $2596.27212 + 2603.95922\,N^2 + 822.999966\,N^4 + 86.8071683\,N^6 + 0.911941440\,N^8$ \\ 
            &   $+ 0.0256180937\,N^{10}$\\
5,1   &   $14967.9450\,N + 7570.43530\,N^3 + 1000.72434\,N^5 + 13.4051039\,N^7 + 0.625521452\,N^9$\\
4,2   &   $21496.0853 + 26926.0889\,N^2 + 6177.06793\,N^4 + 216.477923\,N^6 + 4.80379167\,N^8 $\\
3,3   &   $44213.7572\,N + 17686.3589\,N^3 + 1767.24483\,N^5 + 27.2537306\,N^7$\\
2,4   &   $20802.2607 + 22953.0190\,N^2 + 5124.40912\,N^4 + 171.846652\,N^6 + 0.0681850017\,N^8 $\\
1,5   &   $12267.0624\,N + 4868.45580\,N^3 + 459.187339\,N^5 + 2.19119772\,N^7 $\\
0,6   &   $1359.90833 + 1466.49761\,N^2 + 303.552043\,N^4 + 5.48047785\,N^6 $\\
\hline
7,0   &   $-16956.2524 - 18511.7426\,N^2 - 6735.47522\,N^4 - 924.744197\,N^6 - 32.1576679\,N^8$ \\ 
           &  $+  0.314797775\,N^{10} + 0.0117120916\,N^{12}$   \\
6,1   &   $-119503.736\,N - 69794.4619\,N^3 - 11896.6226\,N^5 - 458.710404\,N^7 + 6.58801236\,N^9$ \\ 
            & $+ 0.363425149\,N^{11}$\\
5,2   &   $-188784.360 - 284537.687\,N^2 - 78459.3804\,N^4 - 4739.14953\,N^6 + 22.5892998\,N^8$ \\ 
            & $+ 3.03196500\,N^{10}$  \\
4,3   &   $-528347.831\,N - 257216.321\,N^3 - 32795.4801\,N^5 - 331.733950\,N^7 + 11.6377106\,N^9 $\\
3,4   &   $-292480.092 - 416299.806\,N^2 - 103998.225\,N^4 - 4515.48789\,N^6 + 8.83297998\,N^8$ \\ 
2,5   &   $-298671.340\,N - 136395.073\,N^3 - 16227.9484\,N^5 - 201.034856\,N^7 - 0.500032757\,N^9 $\\
1,6   &   $-59511.6725 - 77035.0275\,N^2 - 18555.3684\,N^4 - 766.457739\,N^6 - 3.10874638\,N^8$\\
0,7   &   $-13984.0313\,N - 6024.80289\,N^3 - 653.107647\,N^5 - 5.22722079\,N^7 $\\
\hline\hline
\end{tabular}
}

\TABLE[ht]{
\footnotesize
\caption{
The coefficients $b^{v}_{ij}$, cf. Eq.~(\ref{bv}).  
}
\label{betav}
\begin{tabular}{cl}
\hline\hline
\multicolumn{1}{c}{$i,j$}&
\multicolumn{1}{c}{$b^{v}_{ij}$}\\
\hline \hline
4,0   &   0  \\
3,1   &   $234.666985 + 100.520648\,N^2 + 7.96126092\,N^4 - 0.625553655\,N^6$\\
2,2   &   $426.449736\,N + 97.5987129\,N^3 - 2.25343030\,N^5$\\
1,3   &   $254.535978 + 144.854814\,N^2 + 20.3052050\,N^4 $\\
0,4   &   $66.3459284\,N + 18.8271989\,N^3 + 0.560179202\,N^5 $\\
\hline
5,0   &   0  \\
4,1   &   $-1258.17306 - 606.687101\,N^2 - 72.7688516\,N^4 - 1.08252821\,N^6 - 0.287326259\,N^8$\\
3,2   &   $-3271.88287\,N - 894.130891\,N^3 - 25.2461704\,N^5 - 1.55153180\,N^7$\\
2,3   &   $-2559.77598 - 2108.80199\,N^2 - 375.743846\,N^4 - 2.13233677\,N^6 $\\
1,4   &   $-1708.58378\,N - 585.048888\,N^3 - 39.4757354\,N^5 $\\
0,5   &   $-266.102670 - 237.227857\,N^2 - 44.6632567\,N^4 - 0.496927343\,N^6$\\
\hline
6,0   &   0 \\
5,1   &   $8093.66025 + 4354.56520\,N^2 + 652.419979\,N^4 + 13.7782897\,N^6 - 1.54366403\,N^8$ \\ 
            & $- 0.159052164\,N^{10}$\\
4,2   &   $27834.1272\,N + 8658.04010\,N^3 + 371.284973\,N^5 - 18.1481024\,N^7 - 1.18751905\,N^9$\\
3,3   &   $26250.5520 + 27345.7269\,N^2 + 4995.12745\,N^4 - 63.2324448\,N^6 - 3.26781317\,N^8$\\
2,4   &   $30725.1752\,N + 11079.3449\,N^3 + 816.715993\,N^5 - 8.19919537\,N^7$\\
1,5   &   $8470.20827 + 8744.03485\,N^2 + 1815.09720\,N^4 + 39.6191261\,N^6 $\\
0,6   &   $2108.57715\,N + 797.860452\,N^3 + 68.8157914\,N^5 + 0.284187609\,N^7$\\
\hline
7,0   &   0 \\
6,1   &  $-58941.8989 - 34933.0997\,N^2 - 6302.78433\,N^4 - 295.542877\,N^6 + 0.957520387\,N^8$ \\ 
          & $- 1.27213433\,N^{10} - 0.0997279331\,N^{12}$   \\
5,2   &   $-254016.821\,N - 89637.8297\,N^3 - 6227.84400\,N^5 + 32.2770587\,N^7 - 14.8328121\,N^9$ \\ 
        & $- 0.951111299\,N^{11}$                       \\
4,3   &   $-275746.727 - 345193.386\,N^2 - 70372.4321\,N^4 - 538.101215\,N^6 - 67.5278644\,N^8$ \\
 & $- 3.69213913\,N^{10}$  \\
3,4   &   $-476545.568\,N - 197140.916\,N^3 - 19569.5278\,N^5 - 52.0633373\,N^7 - 8.74103639\,N^9$\\
2,5   &   $-174726.233 - 224936.606\,N^2 - 53193.0192\,N^4 - 2001.32523\,N^6 - 7.87773734\,N^8$\\
1,6   &   $-96977.9325\,N - 44001.9263\,N^3 - 5202.93250\,N^5 - 65.8929276\,N^7$\\
0,7   &   $-8282.26037 - 11331.8800\,N^2 - 2764.77952\,N^4 - 113.637842\,N^6 - 0.318785501\,N^8$\\
\hline\hline
\end{tabular}
}

\subsection{Analysis of the series and results}
\label{resultsun}

The knowledge of the large-order behavior of the series allows us to employ  
the conformal-mapping resummation method.
As already mentioned in Sec.~\ref{FDres}, this is essentially
an extension of the method already applied in Refs.~\cite{LZ-77,GZ-98} to
resum the perturbative expansions in O($N$) theories.
Explicitly, given a series
\begin{equation}
R(\bar{u},\bar{v}) = \sum_{hk} R_{hk} \bar{u}^h \bar{v}^k,
\end{equation}
we rewrite it as
\begin{equation}
R(\bar{u},\bar{v}) = \left. R(g \bar{u},g \bar{v}) \right|_{g=1} 
=\hat{R}(\bar{u},\bar{v};g=1),
\end{equation}
where
\begin{equation}
\hat{R}(\bar{u},\bar{v};g) = \sum_n r_n(\bar{u},\bar{v}) g^n,
\qquad r_n(\bar{u},\bar{v}) = \sum_{h+k=n} R_{hk} \bar{u}^h \bar{v}^k,
\end{equation}
Then,
we consider approximants $E({R})_p(\bar{u},\bar{v}; b,\alpha ; g)$
defined by 
\begin{equation}
E({R})_p(\bar{u},\bar{v}; b,\alpha; g) = \sum_{k=0}^p 
  B_k(\bar{u},\bar{v};b,\alpha) 
\int_0^\infty dt\, t^b e^{-t} 
  {Y(g t;\bar{u},\bar{v})^k\over [1 - Y(g t;\bar{u},\bar{v})]^\alpha},
\end{equation}
where
\begin{equation}
Y(x;y,z) = {\sqrt{1 + x A(y,z)} - 1\over 
          \sqrt{1 + x A(y,z)} + 1},
\end{equation}
and $A(y,z)$ is given in Eq.~(\ref{Ares}).
The coefficients 
$B_k(\bar{u},\bar{v};b,\alpha)$ are determined by the requirement that the 
expansion of $E({R})_p(\bar{u},\bar{v};b,\alpha;g)$ in powers of $g$
gives $\hat{R}(\bar{u},\bar{v};g)$ to order $p$. 
For each value of $b$, $\alpha$, and $p$, 
$E({R})_p(\bar{u},\bar{v};b,\alpha;g=1)$
provides an estimate of $R(\bar{u},\bar{v})$.
The parameters $b$ and $\alpha$ are arbitrary and can be
used to optimize the resummation, see, e.g., Refs.~\cite{LZ-77,GZ-98,CPV-00}. 
For example, one may require that the results are least dependent on the 
number $p$ of terms.
Usually one considers
values in the region $0\leq b \lesssim 15$ and  $-1\leq \alpha \lesssim 5$.
The dependence of the results with respect to 
variations of $b$ and $\alpha$ around their optimal values
provides indications of the uncertainty.

We have applied the conformal-mapping
resummation method to the six-loop series
of the $\beta$-functions, and then determined their zeroes.
The zeroes of the $\beta_{\bar{u}}$ and $\beta_{\bar{v}}$ 
are shown in Fig.~\ref{fig} for the cases $N=2,3,4,6$.
In particular, $\beta_{\bar{v}}(\bar{u},\bar{v})$ turns out to vanish
only along the $\bar{v}=0$ axis for any $N$.   
The figures show that no other FP exists
beside the Gaussian and the O($2N^2$) FP's located along the $\bar{v}=0$ axis,
and already predicted by general considerations,
see the discussion at the end of Sec.~\ref{LGWun}.
The $\bar{u}$-coordinate of the O($2N^2$) FP is 
$\bar{u}^*_{O(M)}\approx 1.30,1.19,1.12,1.06$ respectively
for $N=2,3,4,6$  
(these estimates are in agreement with those
that can be found in Refs.~\cite{AS-95,CPRV-96}
for the position of the FP in O($M$)-symmetric theories).
The curves appearing in Fig.~\ref{fig}
are quite robust with respect to the order of the 
series and to variations of the resummation parameters.
None of the approximants of the series for $\beta_{\bar{v}}$, 
from four to six loops,  shows zeroes for $|\bar{v}|>0$ in the region
in which the resummation is still reliable, 
i.e. for $-2\lesssim \bar{u},\bar{v}\lesssim 4$,
which is quite a large region if compared with the location
of the O($2N^2$) FP along the $\bar{v}=0$ axis.
We recall that the relevant region is the one with $v_0>0$
that corresponds to $\bar{v}>0$.
Concerning the zeroes of $\beta_{\bar{u}}(\bar{u},\bar{v})$,
all zeroes obtained by changing the order (from four to six loops)
and by varying $b$ and $\alpha$ in the 
above-mentioned range, lie within the width of the full lines.
Moreover, perfectly consistent results
have also been obtained by employing the Pad\'e-Borel method,
which assumes
only Borel summability, without using the information
on the position of the Borel-transform singularities.
The existence of FP's outside the region 
where the resummation is reliable is quite unlikely.

In conclusion, we do not find evidence of new FP's.
According to RG theory,
this means that a consistent model for a three-dimensional
continuous transition characterized by the symmetry-breaking pattern
$[U(N_f)_L\otimes U(N_f)_R]/U(1)_V\rightarrow U(N_f)_V/U(1)_V$ 
does not exist for any $N\geq 2$. Therefore, the phase transition in such systems
must be a first-order one.
Our results confirm and put on firmer
ground earlier claims \cite{PW-84}
based on a first-order perturbative
calculation within the $\epsilon$-expansion scheme.

\FIGURE[ht]{
\epsfig{file=fig.eps, width=12truecm} 
\caption{
Zeroes of the $\beta$-functions for $N=2,3,4,6$
(solid line for $\beta_u$, dashed line for $\beta_v$).
The grey and black blobs represent the Gaussian and the
O($2N^2$) FP's, respectively.
}
\label{fig}
}

\section{The $SU(N)_L \otimes SU(N)_R$ effective theory}
\label{su4sy}

\subsection{General considerations}
\label{LGWsun}

If the $U(1)_A$ symmetry is explicitly broken by the anomaly,
the axial $U(1)_A$ symmetry is reduced to $\mathbb{Z}(N)_A$
\cite{PW-84}. 
If we rewrite the symmetry-breaking pattern (\ref{symbr}) in the form
$U(1)_A\otimes[SU(N)/\mathbb{Z}(N)]_L\otimes[SU(N)/\mathbb{Z}(N)]_R
\rightarrow SU(N)_V/\mathbb{Z}(N)_V$ and replace $U(1)_A$ with 
$\mathbb{Z}(N)_A$, we obtain 
\begin{equation}
[SU(N)_L\otimes SU(N)_R]/\mathbb{Z}(N)_V\rightarrow SU(N)_V/\mathbb{Z}(N)_V.
\label{symbrs}
\end{equation}
The LGW theory with a complex $N\times N$
matrix order parameter and such a breaking pattern is obtained by adding 
new terms to the Lagrangian ${\cal L}_{U(N)}$. The most relevant one 
is proportional to the determinant of $\Phi$ and thus we obtain the
effective three-dimensional theory 
\begin{equation}
{\cal L}_{SU(N)} = {\cal L}_{U(N)} 
+ w_0 \left( {\rm det} \Phi^\dagger + {\rm det} \Phi \right).
\label{sun}
\end{equation}
Here $w_0$ is related to the breaking of the $U(1)_A$ symmetry.

For $N=1$ the added term is equivalent to
an external field coupled to the order parameter. Therefore, 
it smooths out
the continuous XY transition realized in its absence,
leaving us with an analytic crossover.

For $N=2$ the phase diagram is more complex and is discussed in detail 
in the Appendix. In this case, the effective Lagrangian containing all 
possible interactions with at most four fields is given by 
\begin{eqnarray}
&&
{\cal L}_{SU(2)} = {\rm Tr} (\partial_\mu \Phi^\dagger) (\partial_\mu \Phi)
+r {\rm Tr} \Phi^\dagger \Phi 
\nonumber \\
&&\quad
+ w_0 \left( {\rm det} \Phi^\dagger + {\rm det} \Phi \right)
+ {u_0\over 4} \left( {\rm Tr} \Phi^\dagger \Phi \right)^2
+ {v_0\over 4} {\rm Tr} \left( \Phi^\dagger \Phi \right)^2 
\nonumber \\
&&\quad
+ {x_0\over 4} \left( {\rm Tr} \Phi^\dagger \Phi \right) 
         \left( {\rm det} \Phi^\dagger + {\rm det} \Phi \right) + 
  {y_0\over 4} \left[ ({\rm det} \Phi^\dagger)^2 + ({\rm det} \Phi)^2 \right],
\label{su2} 
\end{eqnarray}
where $\Phi_{ij}$ is a complex 2$\times$2 matrix.

\FIGURE[ht]{
\epsfig{file=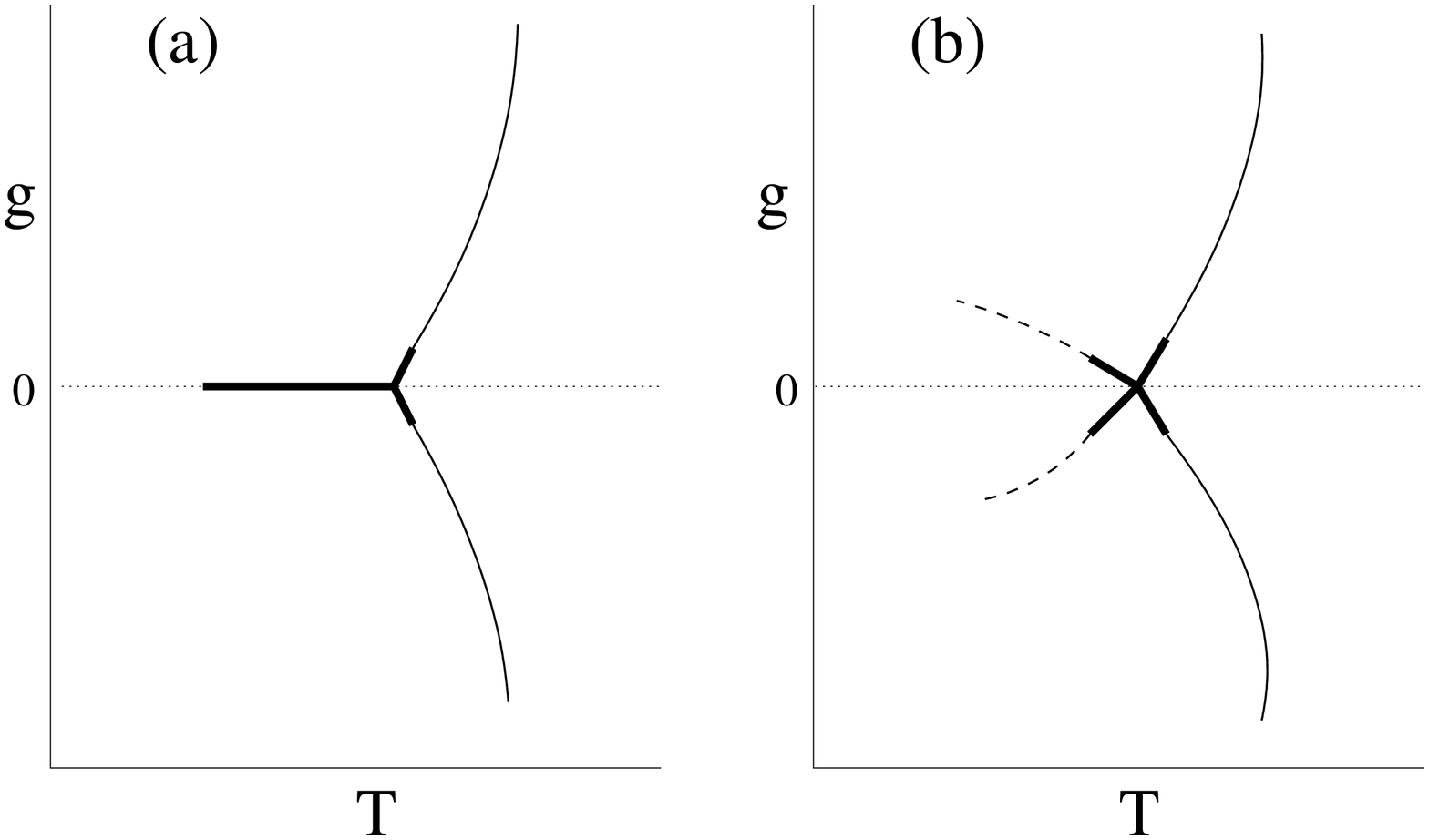, width=12truecm} 
\caption{
Sketches of possible phase diagrams
with bicritical and tetracritical points,
case (a) and (b) respectively.
Full lines correspond to continuous transitions 
with symmetry-breaking pattern $SO(4)\to SO(3)$, 
which are therefore the relevant ones for QCD.
Dashed lines correspond to Ising transitions, not realized in QCD.
Thick lines indicate first-order transitions.
}
\label{phased}
}

Since the Lagrangian (\ref{su2}) has two mass terms, one expects several
transition lines with a multicritical point. 
In the case relevant for QCD the multicritical point
should be identified with the $U(2)\otimes U(2)$
theory, and therefore it should correspond to 
$w_0=0$, $x_0=0$, $y_0=0$. As we discussed in the preceding Section, such 
a multicritical point corresponds to a first-order transition. 
Since a first-order transition is generally robust with respect to 
perturbations, it should persist even for a small breaking of the 
$U(1)_A$ symmetry. Therefore, on the basis of the analysis presented in the
Appendix we expect two possible phase diagrams as a function of the 
temperature $T$ and of the $U(1)_A$ breaking $g$, see Fig.~\ref{phased}. 
In the bicritical case, see Fig.~\ref{phased} (a),
the continuous phase transitions are associated with the
symmetry-breaking pattern (\ref{symbrs}) which is equivalent to 
$SO(4)\rightarrow SO(3)$ \cite{PW-84}.
Therefore, according to universality arguments, if two-flavor QCD 
undergoes a continuous transition, its critical behavior at $T_c$
should be that of the 3-$d$ SO(4) universality 
class \cite{PW-84,Wilczek-92,RW-93},
which has been accurately studied in the literature, see, e.g.,
the recent Refs.~\cite{PV-r,GZ-98,Hasenbusch-00,EM-00,PPV-03}
and references therein.
In the tetracritical case, see Fig.~\ref{phased} (b),
only two continuous transition lines 
are compatible with the symmetry-breaking pattern
expected in QCD, those indicated with the full lines,
see the discussion in the appendix.
Note that the second-order lines end at particular values $g_{\rm tr}$ 
which are tricritical points characterized by mean-field behavior with 
logarithmic corrections.
In conclusion, the transition may be of first-order if the 
$U(1)_A$ breaking is small, second-order in the opposite case, 
and may be of mean-field type for a very specific 
value of $g$.~\footnote{
The effective $U(1)_A$-symmetry breaking
at finite temperature has been investigated on the lattice.
The $U(1)_A$ symmetry appears not to be restored at $T_c$.
However, the effective breaking of the axial $U(1)_A$ symmetry 
appears substantially reduced especially above $T_c$,
as inferred from the difference between the correlators
in the pion and $\delta$ channels.
See, e.g., Refs.~\cite{BCM-97,Betal-97,KLS-98,Vranas-99}.
}
Of course, a specific study of QCD is needed to identify which case 
is effectively realized.  Lattice simulations of two-flavor QCD
favor a continuous transition, see, e.g., 
Refs.~\cite{CP-PACS-01,KLP-01,EHMS-01,KS-01,IKKY-97},
with scaling properties that are substantially consistent with the 
3-$d$ SO(4) universality class.
In the large-$N_c$ limit, $N_c$ being the number of colors,
the anomaly effects are suppressed by powers of $1/N_c$.
Therefore, the finite-temperature transition is expected to be 
of first order.
This picture is also supported by the fact that the finite-temperature
transition in pure SU($N_c$) gauge theories is of first order for $N_c\ge 3$
and in particular in the large-$N_c$ limit, see, e.g., 
Refs.~\cite{LTW-02,Campo-99},
and the quark contributions are expected to be suppressed by 
a factor $1/N_c$ for large $N_c$.
This considerations suggest that for $N_c\ge N_{\rm min}$---Monte Carlo 
simulations indicate $N_{\rm min} > 3$---the finite-temperature transition 
with two flavors changes its nature from continuous to first order.

For $N=3$, the determinant is cubic in the field $\Phi$.
Usually, a cubic term drives to a first-order transition,
and therefore the addition of the determinant is not expected to 
soften the original first-order transition  to a continuous one.
Lattice simulations of QCD confirm this picture,
see, e.g., Refs.~\cite{KLS-01,IKSY-96,Brownetal-90,KS-88,GPS-87}.

For $N=4$ the determinant is a quartic-order term,
giving rise to a generalized LGW  $\Phi^4$ theory with three
quartic parameters.
The effects of the determinant in this case will be discussed 
more carefully in Sec.~\ref{fdexpsu4}.
As we shall see, the determinant does not give rise to 
new FP's, and therefore the phase transition remains
first order also in the presence of a substantial effective
breaking of the $U(1)_A$ symmetry.

Finally, for $N\geq 5$ the determinant term is irrelevant because it
gives rise to polynomial terms of degree higher than four. Therefore,
for $N\geq 5$ the nature of the transition should not change.
The lattice results for $N=6$ of Ref.~\cite{IKSY-96} 
are consistent with this picture.

\subsection{Fixed dimension  expansion in the case $N=4$}
\label{fdexpsu4}

As argued above, the theory for $N=4$ is the only case in which 
it is necessary to investigate the effect of the determinant,
since no arguments exist against the possibility that such a term 
softens the first-order transition expected in its absence.
The three-dimensional RG flow of the Lagrangian ${\cal L}_{SU(N)}$
for $N=4$, cf. Eq.~(\ref{sun}), can be studied by a 
straightforward extension of the
calculations done for the Lagrangian ${\cal L}_{U(N)}$.

Beside the renormalization conditions (\ref{ren1}) and
(\ref{ren2}), we  add a further relation
\begin{eqnarray}
\hat{\Gamma}^{(4)}_{a_1a_2,b_1b_2,c_1c_2,d_1d_2}(0) &=&  
\left. {\delta^4 \Gamma \over \delta \Phi_{a_1a_2}
\delta \Phi_{b_1b_2} \delta \Phi_{c_1c_2} \delta \Phi_{d_1d_2}}\right|_{\rm zero \; mom.}
\nonumber \\
&=& Z_\phi^{-2} m^{4-d} w \epsilon_{a_1b_1c_1d_1} \epsilon_{a_2b_2c_2d_2} ,
\label{ren3}  
\end{eqnarray}
where $\epsilon_{ijkl}$ is the completely antisymmetric tensor ($\epsilon_{1234}=1$).

We computed the perturbative expansion of
correlation functions $\Gamma^{(2)}$, $\Gamma^{(4)}$ and $\hat{\Gamma}^{(4)}$ 
to six loops. Beside the rescaled variables $\bar{u}$ and $\bar{v}$, cf. Eq.~(\ref{rescv})
with $N=4$, we introduce $\bar{w}$ defined by $w = 4 \pi \bar{w}/5$.
The corresponding $\beta$-functions are given by
\begin{eqnarray}
&&\beta_{\bar u}(\bar u,\bar v,\bar w) = 
-\bar u 
+ \bar u^2 + \frac{4}{5} \bar u \bar v + \frac{3}{20} \bar v^2 + \frac{2}{5} \bar w^2
-\frac{751}{5400} \bar u^3 - \frac{4}{27} \bar u^2 \bar v 
\label{bbu} \\
&&- \frac{553}{5400} \bar u \bar v^2
- \frac{41}{225} \bar u \bar w^2 - \frac{1}{25} \bar v^3 - \frac{2}{75} \bar v \bar w^2
+ \sum_{i+j+k \geq 4} \, b^{u}_{ijk} \, \bar u^i \bar v^j \bar w^k ,
\nonumber
\end{eqnarray}
\begin{eqnarray}
&&\beta_{\bar v}(\bar u,\bar v,\bar w) =  
-\bar v 
+\frac{3}{10} \bar u \bar v + \frac{2}{5} \bar v^2  - \frac{2}{5} \bar w^2
-\frac{553}{5400} \bar u^2 \bar v - \frac{4}{27} \bar u \bar v^2 
\label{bbv} \\
&&+ \frac{2}{25} \bar u \bar w^2 
-\frac{73}{1350} \bar v^3 + \frac{2}{45} \bar v \bar w^2
+ \sum_{i+j+k \geq 4} \, b^{v}_{ijk} \, \bar u^i \bar v^j \bar w^k ,
\nonumber 
\end{eqnarray}
\begin{eqnarray}
&&\beta_{\bar w}(\bar u,\bar v,\bar w) =  
-\bar w
+\frac{3}{10} \bar u \bar w -\frac{3}{10} \bar v \bar w
-\frac{553}{5400} \bar u^2 \bar w 
\label{bbw} \\
&&-\frac{11}{1350} \bar u \bar v \bar w 
+ \frac{61}{2700} \bar v^2 \bar w +\frac{4}{225} \bar w^3
+ \sum_{i+j+k\geq 4} \, b^{w}_{ijk} \, \bar u^i \bar v^j \bar w^k. 
\nonumber
\end{eqnarray}
The coefficients 
$b^{u}_{ijk}$, $b^{v}_{ijk}$, and $b^{w}_{ijk}$, with $4\leq i+j+k\leq  7$, 
are reported in the Table~\ref{betas}.
Note that $\beta_{\bar u}$ and $\beta_{\bar v}$ are even functions
of $\bar{w}$, while $\beta_{\bar w}$ is an odd function of $\bar{w}$.

We have also determined the large-order behavior of the series.
Writing
\begin{equation}
S(g\bar u,g\bar v, g\bar w) = \sum_n s_n(\bar u,\bar v,\bar w) g^n,
\end{equation}
we found 
\begin{eqnarray}
&&s_n \sim n! [-A(\bar u,\bar v, \bar w)]^n \, n^b,
\label{bsingw}\\
&&A(\bar u, \bar{v},\bar w) = {6 \pi a \over 5}\,
{\rm Max} \left[ \bar u+\bar v,\, 
\bar u+\frac{\bar v}{4} + \frac{|\bar w|}{2}\right].
\nonumber
\end{eqnarray}
The series turn out to be Borel summable for
\begin{equation}
\bar u+\bar v \geq 0,\qquad
\bar u+\frac{\bar v}{4} - \frac{|\bar w|}{2}\geq 0,
\end{equation}
which is again the stability region 
of the quartic potential (with respect to the bare couplings
$u_0$, $v_0$, and $w_0$).

In order to resum the series of the $\beta$-functions and search for 
their zeroes,
we applied the same method employed in Sec.~\ref{resultsun}.
The analysis of the series clearly shows
that the new $\beta$-function $\beta_{\bar w}$ vanishes
only for $\bar{w}=0$.
As a consequence, the FP's can only lie  in the plane $\bar{w}=0$, and 
therefore they are those of the $U(4)_L\otimes U(4)_R$ theory, see Sec.~\ref{un},
which are unstable.
Since no stable FP is found, the phase transition is expected 
to be first order for any value of $w_0$, and therefore 
also in the presence of a substantial
breaking of the $U(1)_A$ symmetry.

\TABLE[ht]{
\footnotesize
\caption{The coefficients $b^{u}_{ijk}$, $b^{v}_{ijk}$, and $b^{w}_{ijk}$, 
cf. Eqs.~(\ref{bbu}), (\ref{bbv}) and (\ref{bbw}).
Those corresponding to values of $i,j,k$ that are not reported
are zero. 
}
\label{betas}
\renewcommand\arraystretch{0.8}
\begin{tabular}{cllccl}
\hline\hline
\multicolumn{1}{c}{$i,j,k$}&
\multicolumn{1}{c}{$b^{u}_{ijk}$}&
\multicolumn{1}{c}{$b^{v}_{ijk}$}&
\multicolumn{1}{c}{$\qquad\qquad$}&
\multicolumn{1}{c}{$i,j,k$}&
\multicolumn{1}{c}{$b^{w}_{ijk}$}\\
\hline \hline
4,0,0   &   $\phantom{-}$0.0521726542 &  $\phantom{-}$0            &   &   &   \\
3,1,0   &   $\phantom{-}$0.0943284891 &  $\phantom{-}$0.0082425774 &   & 3,0,1   &   $\phantom{-}$0.0082425774 \\
2,2,0   &   $\phantom{-}$0.101307937  &  $\phantom{-}$0.0352787747 &   & 2,1,1   &   $-$0.0286626158 \\
2,0,2   &   $\phantom{-}$0.10529853   &$-$0.0365379374&   &   &    \\
1,3,0   &   $\phantom{-}$0.0612944404 &   $\phantom{-}$0.0485646592&   & 1,2,1   &   $-$0.018209011 \\
1,1,2   &   $\phantom{-}$0.0096560684 &$-$0.00856024106&  & 1,0,3   &   $\phantom{-}$0.0221761335  \\
0,4,0   &   $\phantom{-}$0.013934483  &   $\phantom{-}$0.0127746747 &  & 0,3,1   &   $-$0.000434920883    \\
0,2,2   &  $-$0.033458354&   $\phantom{-}$0.016749908  &  & 0,1,3   &   $\phantom{-}$0.0230423535       \\
0,0,4   &  $-$0.0158081156 & $\phantom{-}$0.00800526372  &   &   &     \\
\hline
5,0,0   &   $-$0.0205830351  &   $\phantom{-}$0           &&   &      \\
4,1,0   &   $-$0.0406573999      &   $-$0.0165182006   &&   4,0,1   &   $-$0.0165182006      \\
3,2,0   &   $-$0.0555154064      &   $-$0.0379950888   &&   3,1,1   &   $-$0.0193283018      \\
3,0,2   &   $-$0.0554550413      &   $\phantom{-}$0.0106667354      &&   &      \\
2,3,0   &   $-$0.0529488118      &   $-$0.0441328387     &&   2,2,1   &   $-$0.01454706        \\
2,1,2   &   $-$0.0101562476      &   $-$0.00163844246    &&   2,0,3   &   $-$0.0155117024      \\
1,4,0   &   $-$0.0250493063      &   $-$0.0264689428     &&   1,3,1   &   $-$0.00480590003     \\
1,2,2   &   $\phantom{-}$0.0285562806       &   $-$0.00933100358    &&   1,1,3   &   $-$0.00990890441     \\
1,0,4   &   $\phantom{-}$0.00460313911      &   $\phantom{-}$0.00264466761       &&   &    \\
0,5,0   &   $-$0.00403223074     &   $-$0.0054784239    &&   0,4,1   &   $-$0.000312387727    \\
0,3,2   &   $\phantom{-}$0.00827027644      &   $-$0.00421798495    &&   0,2,3   &   $\phantom{-}$0.00104252159      \\
0,1,4   &   $-$0.00074739545     &   $\phantom{-}$0.000751462171    &&   0,0,5   &   $\phantom{-}$0.000775960184     \\
\hline
6,0,0   &   $\phantom{-}$0.0108927701       &   $\phantom{-}$0                 &&   &    \\
5,1,0   &   $\phantom{-}$0.0305113637       &   $\phantom{-}$0.000519972577 &&5,0,1 & $\phantom{-}$0.000519972577     \\
4,2,0   &   $\phantom{-}$0.05054934         &   $\phantom{-}$0.00682824039     &&   4,1,1   &   $-$0.00833012972     \\
4,0,2   &   $\phantom{-}$0.0359528478       &   $-$0.00866192578    &&   &    \\
3,3,0   &   $\phantom{-}$0.055702591        &   $\phantom{-}$0.0198339892      &&   3,2,1   &   $-$0.0106000754      \\
3,1,2   &   $\phantom{-}$0.019505026        &   $-$0.00500471528    &&   3,0,3   &   $\phantom{-}$0.0107231224       \\
2,4,0   &   $\phantom{-}$0.0376289338       &   $\phantom{-}$0.0239681302      &&   2,3,1   &   $-$0.00602064077     \\
2,2,2   &   $-$0.00918437911     &   $-$0.00049757924    &&   2,1,3   &   $\phantom{-}$0.00390222416      \\
2,0,4   &   $\phantom{-}$0.00649735532      &   $-$0.00478632298    &   &   &   \\
1,5,0   &   $\phantom{-}$0.0135430912       &   $\phantom{-}$0.0121143686      &&   1,4,1   &   $-$0.00229637061     \\
1,3,2   &   $-$0.00520588781     &   $-$0.00015566449   & &   1,2,3   &   $-$0.00572268276     \\
1,1,4   &   $\phantom{-}$0.00892272331      &   $-$0.00644936705    &&   1,0,5   &   $-$0.00169793557     \\
0,6,0   &   $\phantom{-}$0.00195283172      &   $\phantom{-}$0.00210345121     &&   0,5,1   &   $-$0.000366214748    \\
0,4,2   &   $-$0.000042403606    &   $\phantom{-}$0.0000232000102   &&   0,3,3   &   $-$0.00241486655     \\
0,2,4   &   $\phantom{-}$0.00248127669      &   $-$0.000949407795   &&   0,1,5   &   $-$0.000885328883    \\
0,0,6   &   $-$0.0000138859841   &   $\phantom{-}$0.000230911869    &   &   &      \\
\hline
7,0,0   &   $-$0.00578599786     &   $\phantom{-}$0                 &   &   &      \\
6,1,0   &   $-$0.0167118671      &   $-$0.00498926754    & &  6,0,1   &   $-$0.00498926754     \\
5,2,0   &   $-$0.0309210056      &   $-$0.0159991843     &&   5,1,1   &   $-$0.0108285244      \\
5,0,2   &   $-$0.0223201773      &   $\phantom{-}$0.00295466283     &   &   &     \\
4,3,0    &   $-$0.0426110785      &   $-$0.0268087818     &&   4,2,1   &   $-$0.0134843143      \\
4,1,2   &   $-$0.0163196391      &   $-$0.00135070461    &&   4,0,3   &   $-$0.00802619344     \\
3,4,0   &   $-$0.0402292054      &   $-$0.029458448      &&   3,3,1   &   $-$0.00957047769     \\
3,2,2   &   $\phantom{-}$0.00311298454      &   $-$0.00383770188    &&   3,1,3   &   $-$0.00427874015     \\
3,0,4   &   $-$0.0115432383      &   $\phantom{-}$0.00461784073     &     &   &   \\
2,5,0   &   $-$0.0234111131      &   $-$0.0203943972     &&   2,4,1   &   $-$0.00339634035     \\
2,3,2   &   $\phantom{-}$0.00452594729      &   $-$0.00157893132    &&   2,2,3   &   $\phantom{-}$0.00207216147      \\
2,1,4   &   $-$0.0109265899      &   $\phantom{-}$0.00435906825     &&   2,0,5   &   $-$0.000429888812    \\
1,6,0   &   $-$0.00733233759     &   $-$0.00750892782    &&   1,5,1    &   $-$0.000558386825    \\
1,4,2   &   $\phantom{-}$0.00109452357      &   $-$0.000546285499   &&   1,3,3   &   $\phantom{-}$0.000859221454     \\
1,2,4   &   $-$0.00117934744     &   $\phantom{-}$0.000393239856    &&   1,1,5   &   $-$0.00118266259     \\
1,0,6   &   $\phantom{-}$0.00103714759      &   $-$0.000591465198   &    &   &   \\
0,7,0   &   $-$0.000934334786    &   $-$0.00108103783    &&   0,6,1   &   $-$0.000033642391    \\
0,5,2   &   $\phantom{-}$0.000134905627     &   $-$0.000206454834   &&   0,4,3   &   $-$0.000144490092    \\
0,3,4   &   $\phantom{-}$0.0006868813       &   $-$0.000537767236   &&   0,2,5   &   $-$0.000337875244    \\
0,1,6   &   $\phantom{-}$0.000550816481     &   $-$0.00046918919    &&   0,0,7   &   $-$0.0000195695397   \\
\hline\hline
\end{tabular}
}

\acknowledgments

\noindent
We thank Pasquale Calabrese and Luigi Del Debbio for useful and interesting
discussions.

\appendix

\section{The phase diagram of the 
$SU(2)_L\otimes SU(2)_R$ $\Phi^4$ theory}
\label{appa}

\FIGURE[ht]{
\epsfig{file=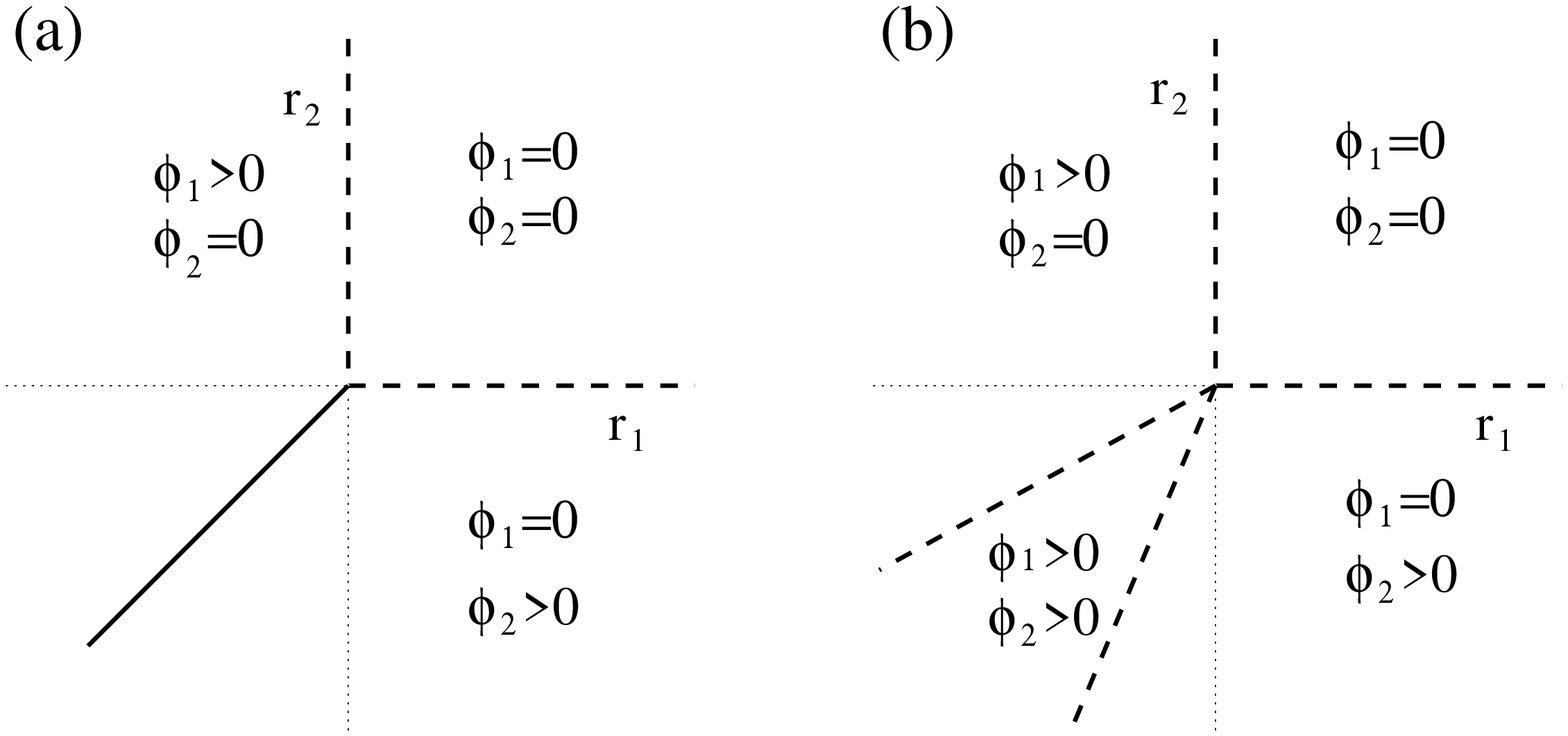, width=12truecm} 
\caption{
Sketches of the possible
phase diagrams in the mean-field approximation:
case (a) is characterized by a bicritical point,
case (b) by  tetracritical point.
The dashed and thick lines represent second- and first-order transitions,
respectively.
}
\label{meanfield}
}

In this appendix we discuss in more detail the phase diagram of 
the theory with Lagrangian (\ref{su2}).
It is easy to map ${\cal L}_{SU(2)}$ into a new one, with two 
4-dimensional vector fields and SO(4) symmetry. If
\begin{equation}
\Phi = \alpha + i \beta + i (\vec{A} + i \vec{B}) \cdot \vec\sigma,
\end{equation}
where $\alpha$, $\beta$, $\vec{A}$, and $\vec{B}$ are real and 
$\vec\sigma$ are the Pauli matrices, we define two fields
\begin{equation}
\phi_1 = 2 (\alpha,\vec{A}), \qquad\qquad
\phi_2 = 2 (\beta,\vec{B}).
\end{equation}
Then Eq.~(\ref{su2}) can be rewritten as 
\begin{eqnarray}
{\cal L}_{SU(2)} &=& {1\over2} (\partial_\mu \phi_1)^2 + 
             {1\over2} (\partial_\mu \phi_2)^2 + 
        {r_1\over2} \phi_1^2 + {r_2\over2} \phi_2^2 
\nonumber \\
&& + {a_0\over 4!} (\phi_1^2)^2 + {b_0\over 4!} (\phi_2^2)^2 + 
     {c_0\over 4!} \phi_1^2 \phi_2^2 + 
     {d_0\over 4!} (\phi_1 \cdot \phi_2)^2,
\label{Lagr-O4}
\end{eqnarray}
with 
\begin{eqnarray}
   r_1 &=& r + w_0, \nonumber  \\
   r_2 &=& r - w_0, \nonumber  \\
   a_0 &=& {3\over4} (2 u_0 + v_0 + 2 x_0 + y_0), \nonumber \\
   b_0 &=& {3\over4} (2 u_0 + v_0 - 2 x_0 + y_0), \nonumber \\
   c_0 &=& {3\over2} (2 u_0 + 3 v_0 - y_0), \nonumber \\
   d_0 &=& - 3 (v_0 + y_0). 
\end{eqnarray}
The Lagrangian (\ref{Lagr-O4}) is stable for $a_0 > 0$, 
$b_0 > 0$, $c_0 > - 2 \sqrt{a_0 b_0}$, and 
$c_0+d_0 > - 2 \sqrt{a_0 b_0}$. It is easy to identify the symmetries of the 
Lagrangian (\ref{Lagr-O4}). It is invariant under O(4) rotations of both 
fields $\phi_1$ and $\phi_2$ and under the $\mathbb{Z}(2)$ transformations 
in which $\phi_1\to -\phi_1$ and $\phi_2\to -\phi_2$ independently
(if the two transformations are applied simultaneously, we obtain an
O(4) transformation).
Therefore, the symmetry group is $\mathbb{Z}(2)\otimes O(4)$. It is interesting
to understand these symmetries in the context of the original model
(\ref{su2}). First, there is the invariance under $\Phi \to A \Phi B$,
$A$, $B\in SU(2)$ that corresponds to the symmetry 
$[SU(2)\otimes SU(2)]/\mathbb{Z}(2)$, which is equivalent to SO(4). 
The other two $\mathbb{Z}(2)$ symmetries correspond to the transformations
$\Phi \to \Phi^\dagger$ and $\Phi\to \Phi^T$.
Note that these two transformations are symmetries of the effective theory
for any $N$. We have not mentioned them before because they do not play any 
role at the QCD transition.

The Lagrangian (\ref{Lagr-O4})
has two independent quadratic (mass) terms and thus it describes 
critical and multicritical transitions.
It is easy to determine the phase diagram of the theory in the 
mean-field approximation, extending the analysis of 
Liu and Fisher \cite{LF-72} . In this case, one assumes that the 
basic fields are space-independent, thereby neglecting the effect of 
fluctuations. 
In this approximations two different phase diagrams are found, 
see Fig.~\ref{meanfield}. In one case, there are three phases: 
1) a disordered one for $r_1>0$ and $r_2 > 0$;
2) an ordered phase in which $\phi_1 \not=0$ and $\phi_2 = 0$, 
with $r_1 < 0$ and $r_2 > \sqrt{b_0/a_0} r_1$;
3) an ordered phase in which $\phi_1 =0$ and $\phi_2 \not= 0$,
with $r_2 < 0$ and $r_2 < \sqrt{b_0/a_0} r_1$.
The multicritical point corresponds to $r_1 = r_2 = 0$, and,
according to standard terminology, is named bicritical in this case.
The phase boundaries between the disordered phase and the ordered ones 
correspond to second-order transitions with symmetry breaking 
$O(4)\to O(3)$, while the boundary $r_2 = \sqrt{b_0/a_0} r_1$, 
$r_1 < 0$, $r_2 < 0$, 
correspond to a first-order transition. 
In the second case, there are four phases: 
1) a disordered one for $r_1>0$ and $r_2 > 0$;
2) an ordered phase in which $\phi_1 \not=0$ and $\phi_2 = 0$,
with $r_1 < 0$, bounded by a line depending on the quartic parameters; 
3) an ordered phase in which $\phi_1 =0$ and $\phi_2 \not= 0$,
with $r_2 < 0$,  bounded by a line depending on the quartic parameters;
4) an ordered phase in which both $\phi_1 \not=0$ and $\phi_2 \not= 0$,
bounded by the two lines considered above.
The multicritical point corresponds to $r_1 = r_2 = 0$, and
is called tetracritical. 
The phase boundaries always correspond to second-order transitions. 

The nature of the phase diagram depends on the values of the quartic parameters.
The multicritical point is bicritical 
for $c_0 > 2 \sqrt{a_0 b_0}$ and $c_0 +d_0 > 2 \sqrt{a_0 b_0}$.
In all other cases it is tetracritical. 
In the tetracritical case, it is of interest to understand the 
symmetry breaking observed at each transition. 
The transitions between the disordered phase and the ordered ones always 
correspond to the breaking $O(4)\to O(3)$, or more precisely to 
$\mathbb{Z}(2)\otimes O(4)\to \mathbb{Z}(2)\otimes O(3)$. 
The additional $\mathbb{Z}(2)$ corresponds to reflections of the 
nonmagnetized field with respect to the plane orthogonal to the 
magnetized one: for instance, if $\phi_1$ is magnetized in the first 
direction, then the symmetry 
$(\phi_2)_1 \to - (\phi_2)_1$, $(\phi_2)_i \to (\phi_2)_i$ ($i\ge 2$) 
is not broken. In the presence of fluctuations these transitions are 
expected to belong to the O(4) universality class.
The transition lines bounding region 4 may correspond to different 
symmetry-breaking patterns. 
If $d_0 < 0$ and 
$- 2 \sqrt{a_0 b_0} < c_0 + d_0 < 2 \sqrt{a_0 b_0}$, 
in phase 4 $\phi_1$ and $\phi_2$ are parallel. 
At the boundaries the $\mathbb{Z}(2)$-symmetry we discussed above is 
broken and the symmetry gets reduced from $\mathbb{Z}(2)\otimes O(3)$ to $O(3)$.
Therefore, these lines should correspond to Ising transitions.
On the other hand,
if $d_0 > 0$ and $- 2 \sqrt{a_0 b_0} < c_0 < 2 \sqrt{a_0 b_0}$ 
in phase 4 $\phi_1$ and $\phi_2$ are orthogonal. 
Therefore, at the transition one observes the symmetry-breaking pattern
$O(3)\to O(2)$ (more precisely 
$\mathbb{Z}(2)\otimes O(3)\to \mathbb{Z}(2)\otimes O(2)$).
In this case, the critical behavior should belong to the O(3) universality 
class.

Finally, let us discuss the phase diagram predicted by 
mean-field theory in the limit in which the effective
breaking of $U(1)_A$ is small. If $g$ parametrizes this breaking,
then $c_0 = 3 u_0 + {9\over2} v_0 + O(g)$, $d_0 = -3 v_0 + O(g)$, and
$\sqrt{a_0 b_0} = {3\over2}(u_0 + v_0/2) + O(g)$, so that 
$c_0 + d_0 = 2 \sqrt{a_0 b_0} + O(g)$. 
Using the results reported above, we obtain that for $v_0 > 0$, i.e. 
$d_0 < 0$, the phase diagram is bicritical or tetracritical depending on the 
sign of the corrections proportional to $g$. Instead, for $v_0 < 0$,
i.e. $d_0 > 0$, the phase diagram is tetracritical.
We recall that the case relevant for QCD should be 
that with $v_0>0$, because only in this case
the symmetry breaking $[U(2)\otimes U(2)]/U(1)\rightarrow U(2)/U(1)$
is realized for $g=0$.

The results we have obtained above in the mean-field approximation
may change when we take into account fluctuations. 
As discussed in Ref.~\cite{KNF-76}, fluctuations change the phase boundaries
that are now generic curves which meet tangentially at the multicritical 
point. It is also possible that some of the transitions become 
of first order. This should happen in the case of interest of QCD.
If $g$ is the anomalous breaking of the axial $U(1)$ symmetry, we know 
that for $g=0$ the transition is of first order. 
Moreover, $g=0$ should be the multicritical point, even in the presence of 
fluctuations, since such a point has a larger symmetry group.
Since a first-order transition is generally robust
under perturbations, the transition should maintain its
first-order nature also for sufficiently
small values of $|g|$, therefore
the transition lines ending at the multicritical point are expected
to be of first order sufficiently close to it.
Possible phase diagrams are sketched in Fig.~\ref{phased}.

\end{document}